\documentclass[aps,twocolumn,showpacs,floats,superscriptaddress]{revtex4}
\usepackage{graphicx}
\usepackage{amssymb,amsmath}
\usepackage{color}
\usepackage{bm}
\begin{document}

\author{Kun Chen}
\affiliation{Department of Physics and Astronomy, Rutgers,
The State University of New Jersey, Piscataway, NJ 08854-8019 USA}
\affiliation{Department of Physics, University of Massachusetts,
Amherst, MA 01003, USA}

\author{Nikolay V. Prokof'ev}
\affiliation{Department of Physics, University of Massachusetts,
Amherst, MA 01003, USA}
\affiliation{National Research Center ``Kurchatov Institute,''
123182 Moscow, Russia}

\author{Boris V. Svistunov}
\affiliation{Department of Physics, University of Massachusetts,
Amherst, MA 01003, USA}
\affiliation{National Research Center ``Kurchatov Institute,''
123182 Moscow, Russia}
\affiliation{Wilczek Quantum Center, School of Physics and Astronomy and T. D. Lee Institute, Shanghai Jiao Tong University, Shanghai 200240, China}

\title{Trapping Collapse}

\date{\today}
\begin{abstract}
Weak potential wells (or traps) in one and two dimensions, and the potential wells slightly
deeper than the critical ones in three dimensions, feature shallow bound states with
localization length much larger than the well radii. We address a simple
fundamental question of how many repulsively interacting bosons can be localized by such
traps. We find that under rather generic conditions, for both weakly and strongly repulsive particles, in two
and three dimensions---but not in one-dimension!---the potential well can trap infinitely many
bosons. For example, even hard-core repulsive interactions
do not prevent this ``trapping collapse" phenomenon from taking place. For the weakly interacting/dilute
regime, the effect can be revealed by the mean-field argument,
while in the case of strong correlations the evidence comes from path-integral simulations.
We also discuss the possibility of having a transition between the infinite and finite
number of trapped particles when strong repulsive inter-particle correlations are increased.
\end{abstract}

\pacs{03.65.-w, 05.30.Jp, 03.65.Ge}

\maketitle
Everyone is familiar with the standard quantum mechanical problem
of finding bound states in a potential well (``trap") of radius $R_0$---for simplicity,
we assume that the potential is spherically symmetric---in one, two, and three dimensions
(1D, 2D, and 3D).
While in 3D the trap strength $V$ has to exceed some finite
critical value $V_c$ to have one bound state, even an arbitrarily weak trap
features a shallow bound state in 1D and 2D; i.e., formally, $V_c=0$ in low dimensions.
For $V-V_c \ll 1/mR_0^2$, with $m$ the particle mass (we will refer to this case as a weak trap),
the only bound state $\psi_1$ is extremely ``shallow":
the binding energy satisfies the condition $ E_1 \ll V-V_c$, and the localization length
 $l= 1/\sqrt{2mE_1}$ (in the units $\hbar=1$) is much larger than the well radius $R_0$.


At the single-particle level, the shallow bound state problem is exhaustively treated in
textbooks (see, e.g., \cite{textbook}). However, to the best of our knowledge, the question of how many
(strongly) repulsive  particles can be localized by a weak trap---and whether the number can be infinite---has
been never addressed. One may immediately deal with two simple cases,
depending on the particle statistics and dimension of space. (i)~A weak
trap cannot bind more than one fermion because even in the absence of
interactions, the second fermion has to go to the delocalized state with zero energy by the
Pauli principle. Thus the best total energy of a pair is $E_2=-E_1$, and adding repulsive
interactions may only increase it further (as a finite-size effect for a delocalized state).
(ii)~A dilute repulsive Bose gas in 1D can be mapped onto non-interacting fermions
(the so-called Tonks-Girardeau limit) when $U > 1/2ml^2 = E_1 $, implying that even for relatively
weak ($U \ll V$) interactions, the trap can bind only one particle, regardless of statistics.
[In this work, we focus on short-range repulsive interactions characterized by a typical interaction
range $R_i \sim R_0$ and potential strength $U$; its zero-momentum Fourier component will be
denoted as $U(0)$ and the $s$-wave scattering length, where appropriate, as $a_s$.]

In all other cases, answering the question requires more elaborate considerations, and, for
strong interactions, numerical simulations. The main result of this work is the effect of trapping collapse when in both 2D and
3D cases, weak traps bind infinitely many bosons. When repulsive interactions are relatively weak
(the criterion is based on the requirement that adding a particle to the system does not change substantially the
structure---and in particular, the density profile---of the state substantially), one can reveal the effect
by solving the Gross-Pitaevskii (GP) equation. Remarkably, the phenomenon holds true even when
interactions are strong, e.g., for hard-core repulsion between bosons, as our lattice path-integral
Monte Carlo (PIMC) simulations confirm. In this case, particles added to the ground state change its structure
substantially. Our data in 2D indicate that in this regime, the localization length diverges
exponentially with the particle number, and we argue that---despite strong correlations---such a behavior can still be
understood based on the GP equation.

The GP approach is justified when many bosons occupy the same mode. For our problem, we
reformulate this condition as a requirement that properties of
the ground state do not change significantly when adding/removing one particle. More specifically,
if a particle is placed into an orbital $\psi$ with the single-particle binding energy $E$ and
the localization length $l_{E}\sim 1/\sqrt{mE}$, the energy of its interaction with the rest of the particles
should remain much smaller than $|E|$. To begin with, this criterion should be satisfied when we
add the second particle to the textbook single-particle state $\psi_1$. In 3D, an estimate
of the potential energy of repulsion immediately follows from properties of bound $s$-wave states
(their asymptotic decay goes as $ \psi_1 = 1/r\sqrt{4\pi l}$):
\begin{equation}
U_i \approx \frac{4\pi a_s}{m} \int d^3r |\psi_1 ({\mathbf r})|^4
\sim \frac{a_s}{ml^2} \int_{R_0} \frac{dr}{r^2} \sim E_1 \frac{a_s}{R_0}.
\label{Upot_3D}
\end{equation}
This leads to the condition for applicability of the GP description of trapped many-particle states
\begin{equation}
a_s \ll R_0 \, \qquad (3D).
\label{condition_3D}
\end{equation}
For short-range interactions with $R_i \sim R_0$, this criterion is no different from the
Born approximation condition for inter-particle scattering.
Since an integral for potential energy is dominated by distances comparable to the smallest scale
in the problem, i.e. the trap radius, this microscopic criterion will not be modified by many-body
effects.

Once  about $\sim R_0/a_s$ particles are placed on the orbital, the $N$-body state will start evolving
towards a more delocalized density profile because the ionization potential defined in terms
of a difference between the total $N$-particle energies as $I_N=E_{N-1}-E_N$ will approach zero.
To see whether the number of localized particles can be infinite, we solve the static
GP equation with zero chemical potential, $\Delta \psi(r) = 8\pi a_s \, \psi^3$, or
\begin{equation}
\psi''(r)+2\psi'(r)/r = 8\pi a_s \, \psi^3 \,.
\label{GP_3D}
\end{equation}
The solution decays at large distances $\psi(r\to \infty) = 1/\sqrt{16\pi a_s \ln(r/R_0)}\, r$,
but this decay is weak:  It does not prevent the integral for the total particle number,
$N=\int d^3r |\psi (r)|^2 \propto \int dr/\ln(r/R_0)$, from diverging at the upper limit.
This establishes the effect of trapping collapse for a weakly interacting 3D system.
As for the total interaction energy, the integral $\int d^3r |\psi (r)|^4$ is still dominated
by the trap potential region.

Similar considerations apply to the 2D case with one notable exception: the effective
repulsive interaction is now scale-dependent with logarithmic renormalizion
towards smaller value at low energies:
\begin{equation}
U_{{\rm eff}} (k) \approx \frac{U(0)}{1 + g\ln (1/kR_i)} \, , \qquad g=\frac{mU(0)}{2\pi} \, ,
\label{Ueff}
\end{equation}
where $k$ is the relative momentum of two particles. This formula, in particular, implies that even
for strongly repulsive bosons with $g \gg 1$, the effective interaction is weak (and universal!):
$ m U_{{\rm eff}} (k) \to 2\pi / \ln (1/kR_i) \ll 1$ at low enough energies.
An estimate of the potential energy of repulsion for two particles now reads:
\begin{equation}
U_{{\rm i}} \approx  U_{{\rm eff}} (1/l)  \int d^2r |\psi_1 ({\mathbf r})|^4
\sim E_1 m U_{{\rm eff}} (1/l) \,.
\label{Upot_2D}
\end{equation}
The integral is dominated by distances of the order $l$, justifying the use of the
effective coupling constant (recall that in 2D the bound state described by the modified
Bessel function $K_0(r/l)$ is only weakly dependent on distance under the localization length).
If the interaction is weak, $mU(0)\ll 1$, or the state is very shallow, we find that conditions for
applying the GP equation to the trapping problem are satisfied. The solution of the radial
equation at zero chemical potential,
\begin{equation}
\psi''(r)+\psi'/r = 2 m U_{{\rm eff}}(r) \, \psi^3 \, ,
%
%
\label{GP_2D}
\end{equation}
has the asymptotic form $ \psi(r \to \infty ) =\sqrt{\ln (r/R_0) /4\pi }/r $.
Here we explicitly consider the universal asymptotic expression for $m U_{\rm eff}$. If the
value of the constant $g$ in Eq.~(\ref{Ueff}) is small, and the logarithmic flow of the coupling
constant can be neglected, then in a broad range of intermediate length scales the solution
is simply $ \psi(r) = 1 / \sqrt{4\pi g}\, r $.
Again, the integral for the total particle number diverges at the upper limit,
$N \propto \int dr \ln(r/R_0) /r $, indicating that weak traps in 2D can localize infinitely
many bosons, including strongly repulsive ones. [Logarithmic dependence of $N$ on the upper cutoff
ensures that finite-density corrections to the $U_{{\rm eff}}(r)$ dependence on length scale
remain sub-leading.]

To verify these results, and to explore what happens when conditions for the GP approach
are not satisfied in the form of strong inequalities, we resort to PIMC simulations of
square/cubic lattice systems with tight-binding dispersion relation
$\epsilon({\mathbf k}) = 2 t \sum_{\alpha=1}^d [1-\cos (k_\alpha a)]$,
where $t$ is the nearest-neighbor hopping matrix element and
$a$ is the lattice constant (in what follows, the energy and distance are measured in units of $t$ and $a$, respetively).
The trap is introduced as an attractive potential $-V \delta_{{\mathbf r},0}$ placed at the origin.
The repulsive pairwise interactions between particles are of the on-site Hubbard form,
$U({\mathbf r}_i-{\mathbf r}_j) =U\delta_{{\mathbf r}_i-{\mathbf r}_j, 0}$.
In 2D, our study of localized many-particle states was performed for $V=2$ to ensure that
the binding energy is about a factor of one hundred smaller than the bandwidth, $E_1=0.0576$.
The critical value of $V$ for forming a bound state in a cubic lattice is $V_c=3.956776$;
for the trapping  potential strength $V=4.3$ used in this work the binding energy is only
$E_1 = 0.06058$.

\begin{figure}
\centering
\includegraphics[width=0.9\linewidth]{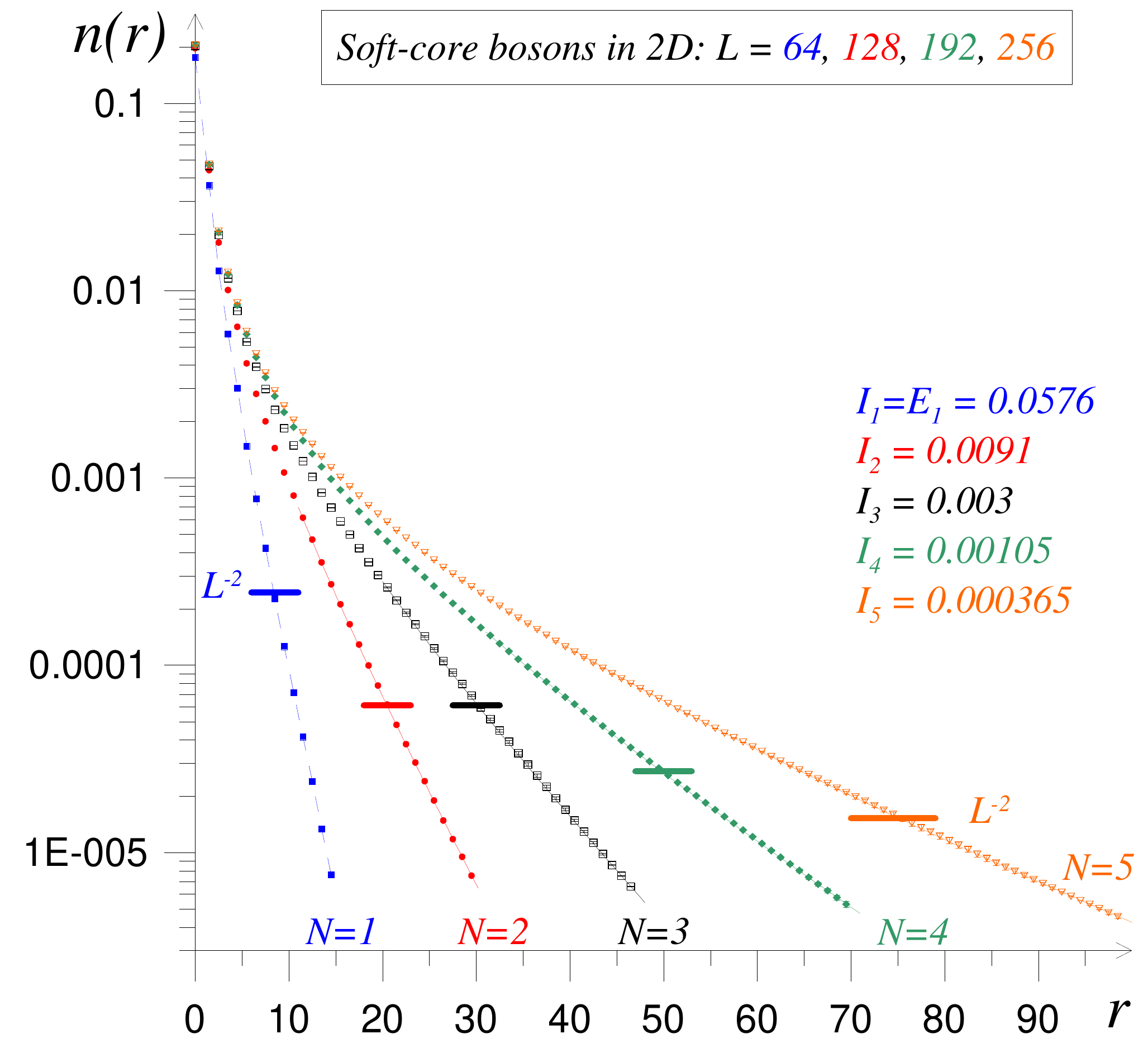}
\caption{\label{fig:1} (color online).
Radial density profiles in 2D for $V=-2$ and $U=4$ at $\beta = 16000$. System sizes
were increased for larger $N$ to ensure that $n(r)$ for weakly bound states drops well below the
$L^{-2}$ level (shown by the bold bar) characteristic of a delocalized single-particle state.
Error bars are smaller than symbol sizes unless shown explicitly. Solid fitting lines are
explained in the text.
}
\end{figure}
\begin{figure}
\centering
\includegraphics[width=0.9\linewidth]{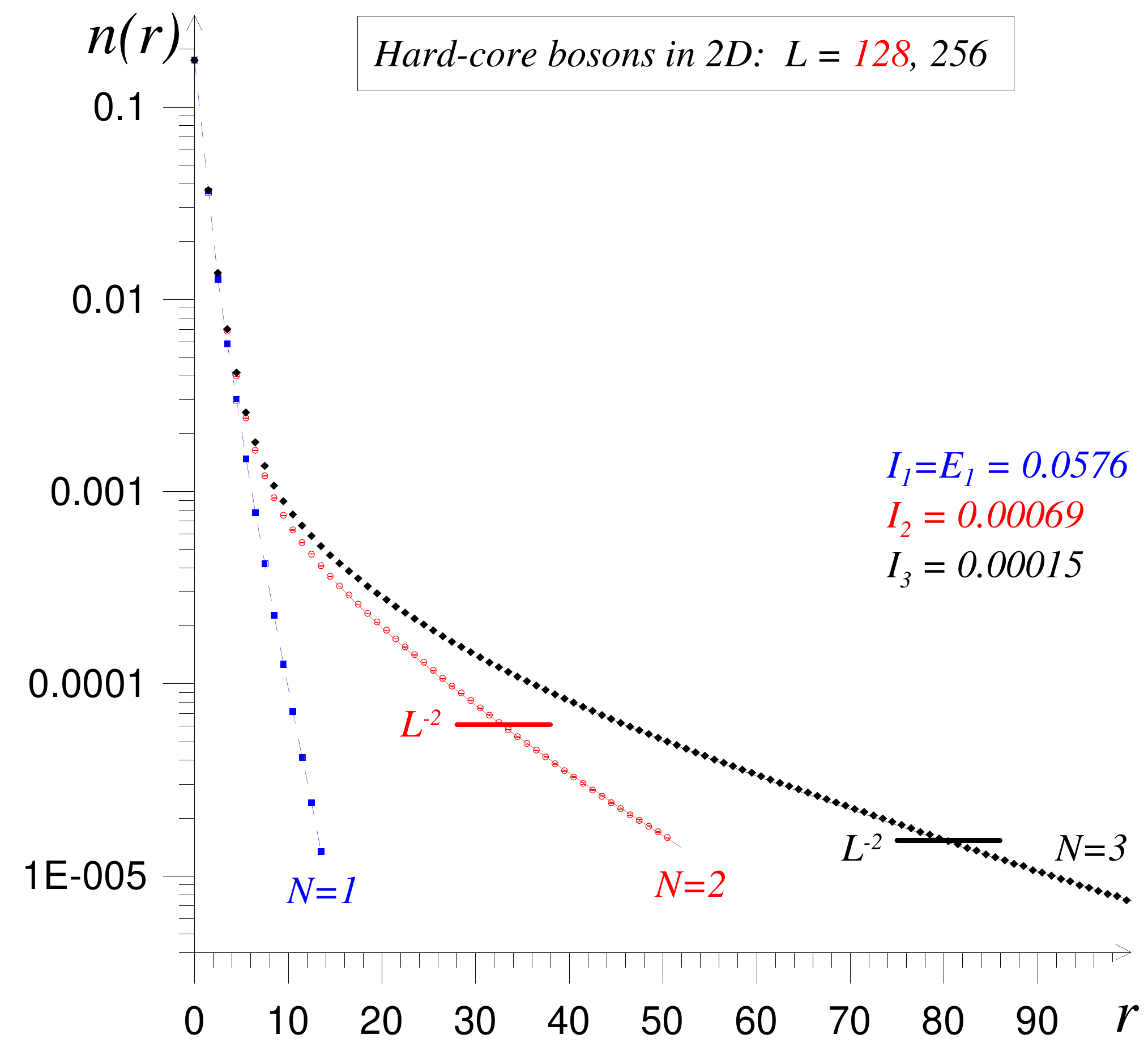}
\caption{\label{fig:2} (color online).
Radial density profiles for hard-core bosons in 2D for $V=-2$ at $\beta = 16000$
(see Fig.~\ref{fig:1} caption for additional details that are identical for both figures).
}
\end{figure}
\begin{figure}
\centering
\includegraphics[width=0.9\linewidth]{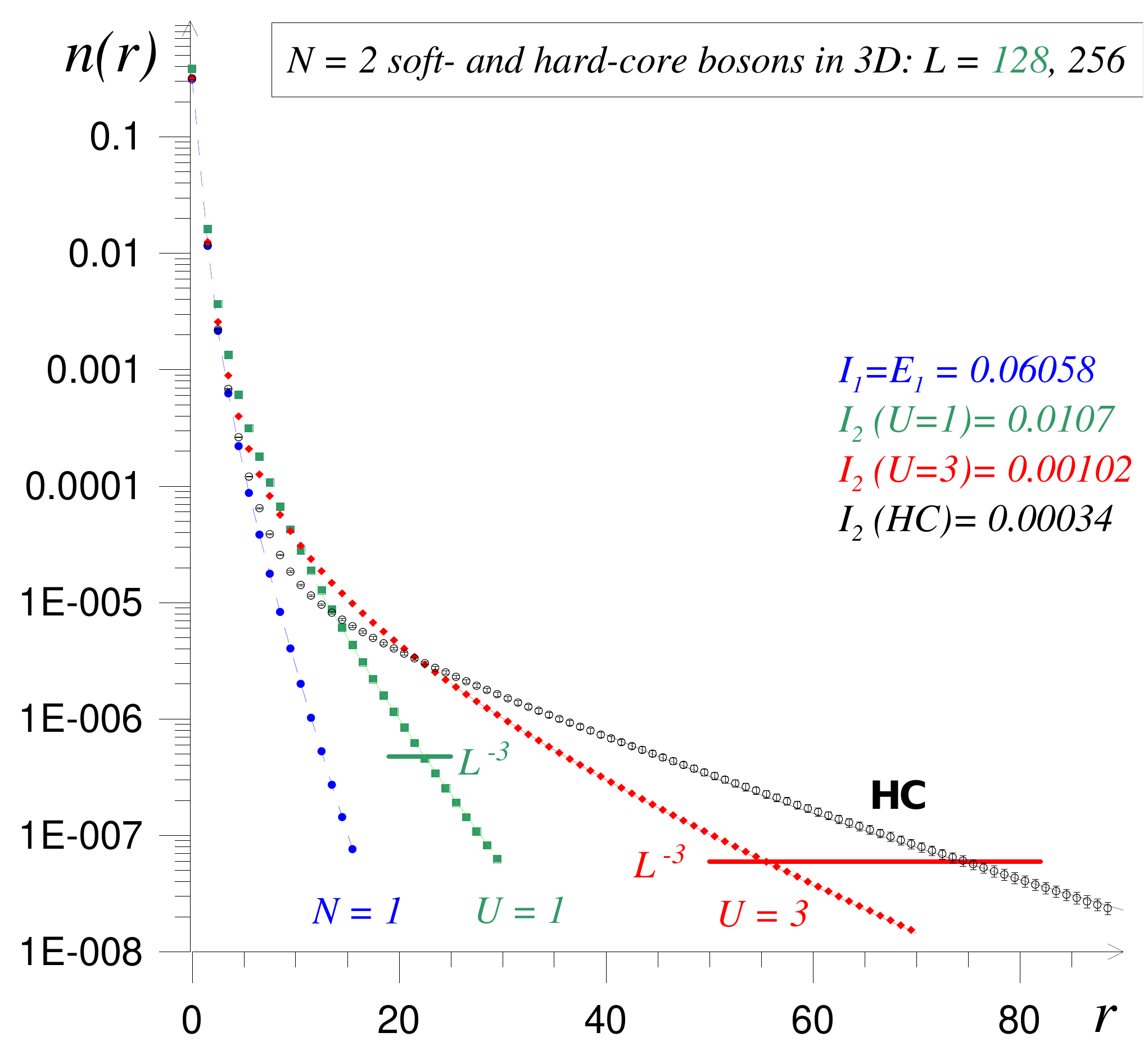}
\caption{\label{fig:3} (color online).
Radial density profiles for two soft- and hard-core bosons in 3D for $V=-4.3$ at $\beta = 10000$.
The $L^{-3}$ level characteristic of the delocalized single-particle state is shown
by bold bars. Large scale decays are fitted to the $e^{-2r/l_2}/r^2$ law.
}
\end{figure}
\begin{figure}
\centering
\includegraphics[width=0.9\linewidth]{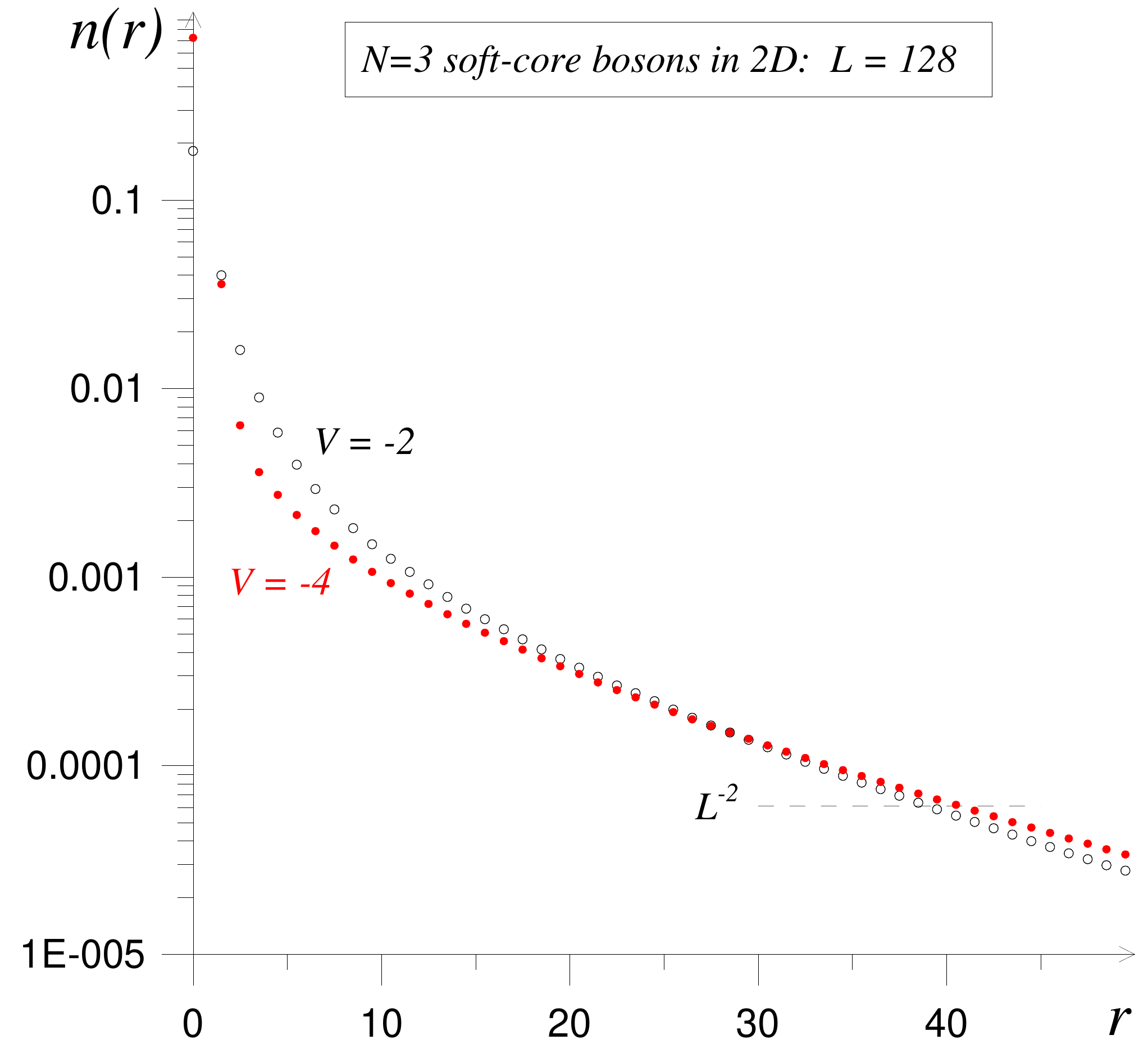}
\caption{\label{fig:4} (color online).
Radial density profiles for three bosons in 2D for $U=10$ in traps with $V=2$ and $V=4$
at $\beta = 4000$, showing that the state is more localized in a weaker trap.
}
\end{figure}

In Figs.~\ref{fig:1} -- \ref{fig:4}, we show data for localized density profiles of multi-particle states
in weak traps. The data are averaged over circular/spherical bins of unit length in the radial direction.
With the Monte Carlo algorithm optimized for simulations of dilute systems, when hundreds of kinks
are changed in a single elementary update, we were
able to address ground state properties of very large systems (in all fixed-$N$ simulations
the inverse temperature $\beta =1/T$ was large enough to guarantee that contributions from
excited states were negligible).
As the localization length rapidly increases with the number of particles, we ultimately
hit the computational complexity threshold at some finite $N$. In 2D, for inter-particle repulsion
strength $U=4$, we were able to quantify properties of localized states of up to five bosons,
see Fig.~\ref{fig:1}. Clearly, having the inter-particle repulsion a factor of two stronger than
the trap potential does not stop the system from forming a localized many-body ground state.

By fitting profile tails to the asymptotic decay of the $K_0^2(r/l_N)$ function,
we deduce the ionization potential of the $N$-particle state from $I_N=1/2ml_N^2$.
It is evident that $I_N$ quickly diminishes with $N$ (this is the prime
reason for why we need large systems and extremely low temperatures to reveal localized states).
For $N\ge 2$ the ionization potentials can be fitted well by an exponential
function $I_N \propto e^{-cN}$ with constant $c$ close to unity.
This result is consistent with the mean-field picture described by the GP equation.
For $U=4$, the bare coupling parameter $g=1/\pi $ is smaller than unity, and the GP solution at relevant scales
decays as a power law $ \psi(r) = 1 / 4\pi g r $. This leads to an approximate
relation between the particle number and localization length, $N \sim (1/2g) \ln(l_N)$,
that can be used to estimate the ionization potential $ \ln(I_N) = -\ln(2ml_N^2) \propto -N/4g$.

Trapping collapse phenomenon persists even when the on-site repulsion is taken to the ultimate
hard-core (HC) limit, see Fig.~\ref{fig:2}. In this case, we are certainly not in the GP regime
for $N=2,3$ since the density profile undergoes radical changes by adding one particle. Rather,
we are dealing with a strongly correlated state such that when one particle is within
the localization length $l$ from the trap center, the other particles are most likely to be
found at a much larger distance, leading to a bi-modal structure of $n(r)$. For $N=3$ the
effects of strong correlations are pronounced, but are less dramatic quantitatively than for $N=2$.
The divergence of the localization length with $N$ in this case is much faster than for
$U=4$, and for $N=3$ the ionization potential drops down to $0.00015$, limiting our ability
to monitor the crossover to the GP picture. However, given precise understanding of what
happens in dilute 2D systems at large scales, there is little doubt
that at zero chemical potential the ground state involves infinitely many HC particles.

We find similar results for 3D systems, see Fig.~\ref{fig:3}.
For $U=1$ (relatively weak coupling) the two-particle bound state
resembles that of two bosons being placed on the same orbital, but even then the ionization potential
is about a factor of six smaller than $E_1$. For $U=3$ we are
already dealing with the ground state where positions of two particles are strongly correlated.
For hard-core bosons, we certainly violate the condition (\ref{condition_3D}), and the
two-body state develops a signature bi-model shape when at short distance the two-body
density profile is closely following a single particle one, see Fig.~\ref{fig:3}.
This appears to be the generic mechanism for particles to minimize effects of strong
repulsive interactions while gaining enough potential energy from the trap to remain
in the localized state. The energy balance, however is extremely delicate: the ionization
potential $I_2(HC)$ is nearly two hundred times smaller then $E_1$!

\begin{figure}
\centering
\includegraphics[width=0.9\linewidth]{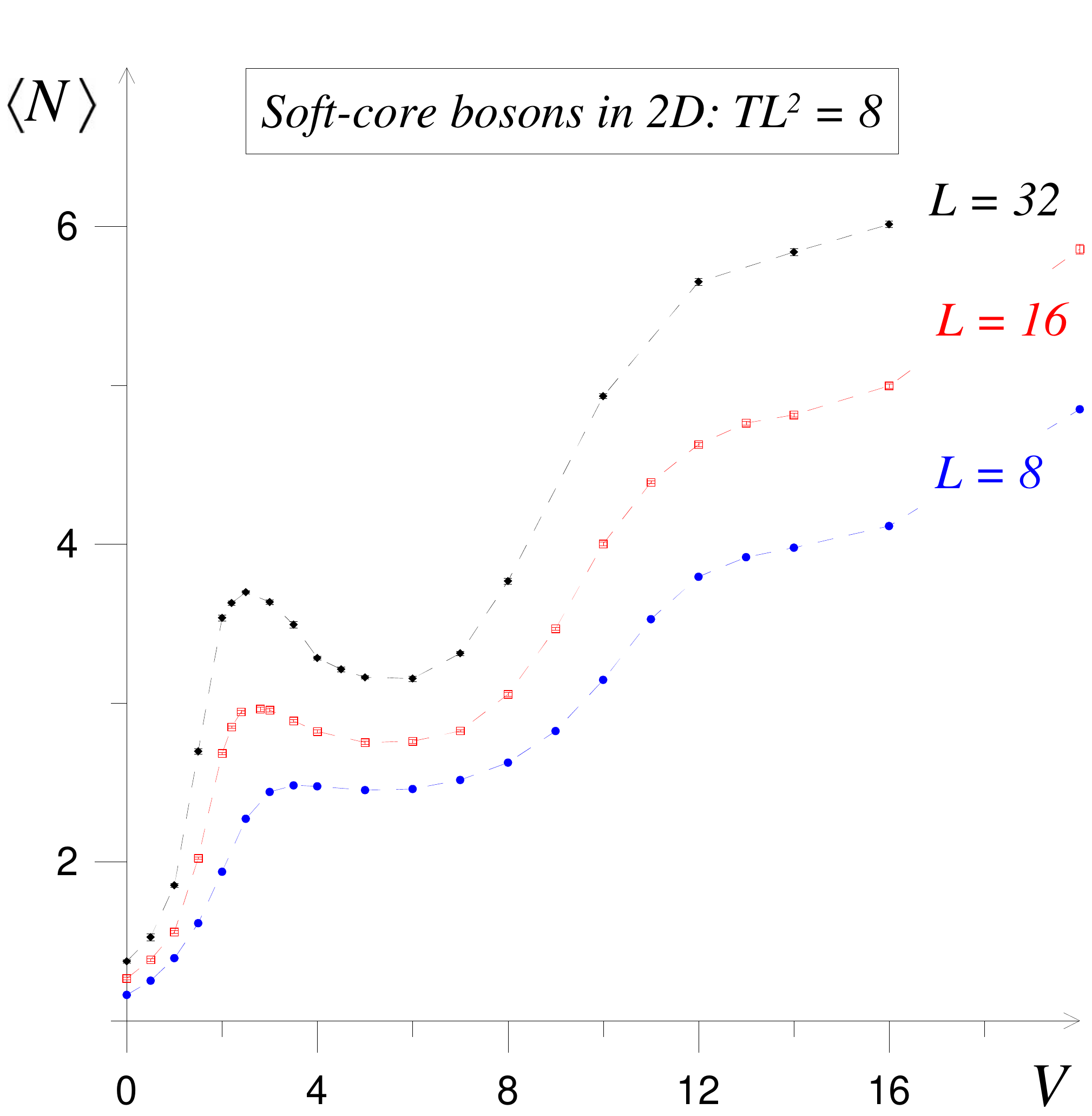}
\caption{\label{fig:5} (color online).
Average number of soft-core bosons with $U=10$  at zero chemical potential as a function of trap strength
for different system sizes and temperatures in 2D.
}
\end{figure}

The most intriguing question that remains unanswered by the data, is the transition
between the ground state with infinitely many localized particles and a state
with a few, or just one, localized particles as the range and strength of the repulsive
interaction is increased, or, counter-intuitively, when the trap potential is increased.
Indeed, according to expression (\ref{Ueff}), effects of repulsive interactions are
more pronounced for smaller localization length $l$, or deeper traps. Strong-correlation
effects then result in a state where one particle stays close to the trap center and
effectively ``screens" it out. This effect is verified and quantified
in Figs.~\ref{fig:4} and \ref{fig:5} using two different setups.
In Fig.~\ref{fig:4}, we directly observe that the state of three bosons with strong on-site
repulsion $U=10$ is more delocalized in a deeper trap with $V=4$ than in a trap with the
shallow single-particle state at $V=2$.

Higher energies for multi-particle states in a deeper trap imply that the average
particle number at a given temperature and zero chemical potential must have a
minimum at some value of $V$ (when $V$ is larger than $U$,
one tightly localized particle can no longer fully screen the trap). Minima on the $\langle N \rangle$ curves
as a function of $V$ for 2D soft-core boson with $U=10$ are clearly seen in Fig.~\ref{fig:5}.
As far as evidence goes, increasing trap potential for strongly repulsive bosons with $U=10$
does not lead to the trapping collapse transition: we do not observe saturation of the
$\langle N \rangle$ curves to some finite thermodynamic limit answer when the system size
is increased (temperature is decreased accordingly to keep the product $TL^2=8$ fixed),
see Fig.~\ref{fig:5}.

In conclusion, we have found that two- and three-dimensional finite-range potential wells,
including those featuring only one weakly bound single-particle state, will localize
infinitely many bosons even when repulsive inter-particle interactions are much stronger
than the trapping potential. We termed this effect the trapping collapse,
tracing its origin in the mean-field regime captured by the Gross-Pitaevskii equation.
Evidence for trapping collapse in the case of strong repulsive interactions was provided
by path-integral Monte Carlo simulations. Future work should clarify under what conditions the
trapping collapse phenomenon is replaced with localization of a finite (one ?) number of
particles and what are properties of systems at the transition point.

This work was supported by the National Science Foundation under the grant DMR-1720465
and the MURI Program ``Advanced quantum materials -- a new frontier for ultracold atoms" from AFOSR.

\end{document}